# Automating Cobb Angle Measurement for Adolescent Idiopathic Scoliosis using Instance Segmentation

Chaojun Chen[1]
Khashayar Namdar[2,3,6]
Yujie Wu[1]
Shahob Hosseinpour[4]
Manohar Shroff[4]
Andrea S. Doria[4]
Farzad Khalvati[1,2,3,4,5,6]

October 2022

1. Department of Mechanical and Industrial Engineering, University of Toronto, Toronto, ON, Canada
2. Institute of Medical Science, University of Toronto, Toronto, ON, Canada
3. Department of Diagnostic Imaging, Research Institute, The Hospital for Sick Children, Toronto, ON, Canada
4. Department of Medical Imaging, University of Toronto, Toronto, ON, Canada
5. Department of Computer Science, University of Toronto, Toronto, ON, Canada
6. Vector Institute, Toronto, ON, Canada

## Abstract

Scoliosis is a three-dimensional deformity of the spine, most often diagnosed in childhood. It affects 2-3% of the population, which is approximately seven million people in North America. Currently, the reference standard for assessing scoliosis is based on the manual assignment of Cobb angles at the site of the curvature center. This manual process is time consuming and unreliable as it is affected by inter- and intra-observer variance. To overcome these inaccuracies, machine learning (ML) methods can be used to automate the Cobb angle measurement process. This paper proposes to address the Cobb angle measurement task using YOLACT, an instance segmentation model. The proposed method first segments the vertebrae in an X-Ray image using YOLACT, then it tracks the important landmarks using the minimum bounding box approach. Lastly, the extracted landmarks are used to calculate the corresponding Cobb angles. The model achieved a Symmetric Mean Absolute Percentage Error (SMAPE) score of 10.76%, demonstrating the reliability of this process in both vertebra localization and Cobb angle measurement.

# 1 Introduction

Scoliosis is an abnormal lateral curvature of the spine, often diagnosed in childhood or early adolescence. It causes chronic back pain as well as uneven shoulder and waist [1]. Severe scoliosis can be disabling. An especially severe spinal curve requires surgical intervention as it reduces the amount of space within the chest, making it difficult for the lungs to function properly [2]. Therefore, it is important to identify scoliosis at an early stage and provide appropriate treatment by the time the condition can be reverted or improved with braces rather than by the time it requires corrective surgery.

The diagnosis and severity of scoliosis are quantified by assessing the Cobb angles, which are commonly measured using anterior-posterior (AP) radiography (X-ray). Due to the ambiguity and variability in the scoliosis AP X-ray images, measuring Cobb angles is a challenging task [4]. Generally, radiologists assess Cobb angles by manually selecting the most tilted vertebrae above and below the apex, as shown in Figure 1. Lines are then drawn along the endplates of each vertebra, and the angle between the lines is the Cobb angle [5]. This common process can be affected by two main sources of error, in addition to other anatomic conditions such as segmentation spinal anomalies. The first one is inter-observer variance, which refers to the difference in end vertebra selection by different radiologists. The other error is intra-observer variance, which captures the fact that a radiologist might identify different end vertebra on the same x-ray image at different times [6]. These differences can lead to inaccuracy and inconsistency in Cobb angle measurement [6].

To correct for these inaccuracies, previous research proposed to automate Cobb angle measurements with machine learning (ML) approaches [7]. Key components in solving this problem include: (1) correct localization of the vertebrae, (2) correct extraction of landmarks, and (3) correct calculation of Cobb angles. To do this, one common approach is to first use supervised learning to identify the location of each vertebra; then use a computer vision algorithm to extract the important landmarks, which are the four corners of each vertebra; and lastly utilize the extracted landmarks to calculate Cobb angles through trigonometry [8]. One way to improve the accuracy of this method is to use a more modern segmentation model for the vertebra localization task. Thus, the objective of this research is to create a consistent, accurate, and autonomous solution for Cobb angle measurement through ML which will help clinicians develop optimal treatment for patients by minimizing errors.

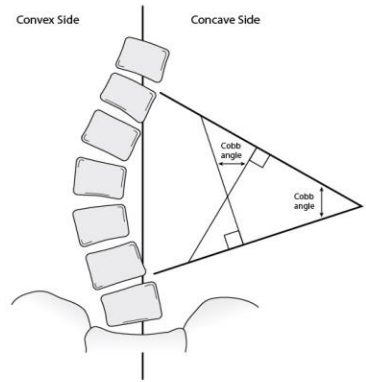

**Figure 1:** Cobb angle measurement procedure shows how the angle of spinal curvature is measured by drawing lines parallel to the upper border of the upper vertebral body and the lower border of the lowest vertebra of the structural curve. The angle that is generated when perpendiculars from these lines cross each other constitutes the Cobb angle[3].

## 2 Previous Work

### 2.1 Spine Localization

Automating Cobb angle measurement involves the correct localization of the spine and vertebrae. Some of the methods for this task are regression-based method, segmentation-based method, and object detection-based method.

#### 2.1.1 Regression-based methods

Regression-based methods estimate the landmark locations and Cobb angles directly. One technique is to jointly evaluate Cobb angle and landmarks of the spine using Structured Supported Vector Regression ($S^2VR$) [9]. The nonlinear mapping process and the explicit correlation learning stage within $S^2VR$ allow it to perform regression tasks based on the extracted image features. Sun et al. achieved a Relative Root Mean Squared Error (RRMSE) of 21.63% on a dataset from London Health Sciences Center [9]. Similarly, BoostNet, a model that leverages the robust feature extraction capabilities of Convolutional Neural Networks (CNN), uses regression to determine landmarks [10]. Wu et al. used Mean Square Error (MSE) to measure the performance of BoostNet and achieved an MSE of 0.0046 on the test dataset [10]. Regression-based methods can capture the global information of the images. However, they require a significant number of parameters due to dense mapping between the regressed points, leading to high computational costs [9]. As a result, the input images which can be as high resolution as 2600×1600 pixels, have to be shrunk to a much smaller resolution (256×128 pixels) for training and inference [11]. This down sampling limits the performance of the models, as detailed information is lost during this process.

#### 2.1.2 Segmentation-based methods

One way to address the issues of the regression-based methods is to utilize a segmentation model for vertebrae localization. Popular models such as U-Net and Mask RCNN were used in previous research [12]. U-Net is a deep model that performs semantic segmentation and labels each pixel of an image with a corresponding class [13]. In this case, the convolutional layers inside U-Net help extract important features in the image and identify the pixel pattern of each vertebra. Horng et al. utilized a revised U-Net on a dataset of 595 vertebra images and achieved an MSE of 0.025 on the test cohort [14]. Another type of segmentation model used previously is Mask RCNN. Pan et al. applied two Mask RCNN models to detect spine and vertebral bodies separately. The midpoints of the endplates of each vertebra were then determined from the output masks. The paper uses a dataset of 248 chest x-ray images from Ruijin Hospital in Shanghai, China and the patients were between 22 and 93 years of age. This method achieved a Mean Absolute Difference (MAD) of 3.32 degrees, demonstrating the models' reliability [15]. In general, segmentation method provides acceptable results; however, it is sensitive to image qualities. Image preprocessing is required in some cases to remove image noise, enhance image resolution, and adjust image contrast [16]. It was noted in previous research that when input images have low image quality, segmentation models tend to predict connected or corrupted vertebra masks.

### 2.1.3 Object-detection-based methods

Object-detection-based methods can be used to solve the corrupted segmentation mask problem described above. Khanal et al. proposed a novel approach to first detect 17 vertebrae with a bounding box object detector, then pass the predicted box to a densely connected CNN to find the four corners of each vertebra [17]. By first detecting vertebrae as an object using Faster RCNN, the authors reduce the search space for landmarks detector, leading to more accurate results. However, it is worth noting that this approach does not consider the inter-dependency between landmark positions. Faster RCNN achieves a calculated Symmetric Mean Absolute Percentage Error (SMAPE) of 25.69% on a dataset of 609 images [17]. In another research, Yi et al. proposed to detect the center of each vertebra then trace the four corners through the learned corner offset using ResNet [11]. By first detecting the center, the algorithm can keep the landmarks in order and avoid the overlapping issue that arises in the segmentation method. This approach considers the vertebra's inter-dependency and achieves an average SMAPE of 15.9% on a dataset of 609 images [11].

## 3 Method

Aiming to select an accurate, consistent and optimal approach that does not require an excessive amount of computation resources, we chose a segmentation-based method. Based on the literature, segmentation-based methods generally have a lower SMAPE score. Additionally, they are able to intake higher-resolution input images and localize points without dense mapping. Therefore, this paper will approach the automation task using YOLACT segmentation method [18].

### 3.1 Segmentation using YOLACT

YOLACT is an instance segmentation with a ResNet backbone that produces a feature pyramid based on the input image [18]. The information inside the feature pyramid is then passed into two parallel pathways. One pathway is the Protonet which has four convolution layers and performs the segmentation task. The second pathway finds the mask coefficients of the image based on RetinaNet, which produces image masks and bounding boxes. Finally, the algorithm linearly combines the two outputs and uses sigmoid nonlinearity to produce the final mask.

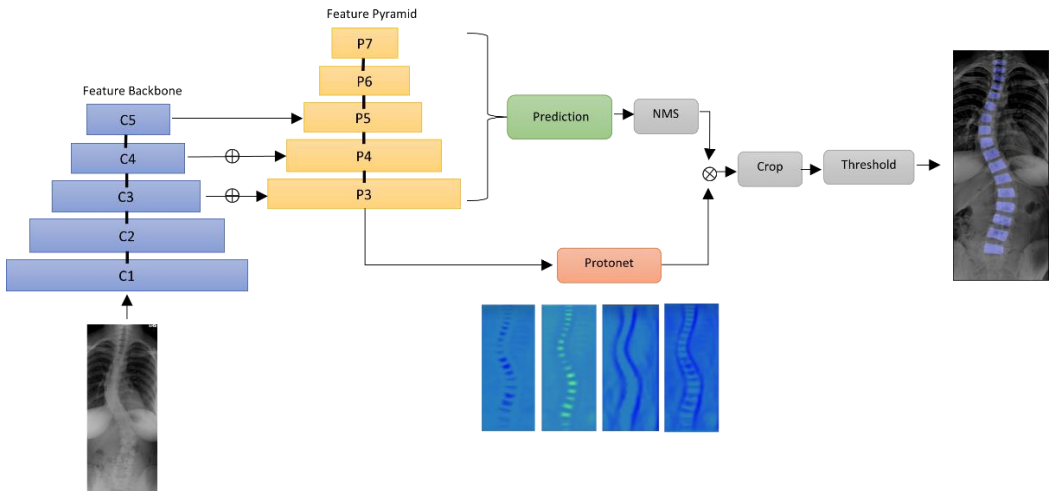

**Figure 2:** YOLACT Architecture is based on RetinaNet, ResNet-101, and Feature Pyramid Network (FPN) [18]. Information produced by the Feature Pyramid will is passed into two parallel pathways, Prediction Head and Protonet, for further evaluation

### 3.1.1   Protonet

The protonet is implemented as a Fully Convolutional Network (FCN), and it is attached to the deepest feature layer to produce robust and high quality masks. The arrows in 3 represent 3×3 convolution layers, except for the last layer that is a 1 x 1 convolution layer. The increase in size is the result of an up-sample followed by convolution layer. The last layer has *k* channels, one for each prototype [18]. The prototypes are candidate masks and they capture spacial information. In this case, the number of prototypes is set to 32 (ie. k=32) to achieve the optimal mix of speed and performance. Some examples of the prototypes are shown in Figure 4. This formulation is slightly different from the standard semantic segmentation as there is no explicit loss on the prototypes. All supervision of the prototypes happens after mask assembly.

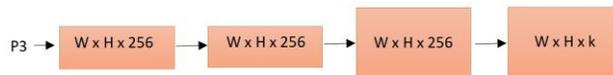

**Figure 3:** Protonet Architecture is attached to the deepest and largest feature layers of the FPN. Protonet takes in images of size W x H and each arrow represents a 3 x 3 convolution layer except for the final layer, which is 1 x 1. The final output of Protonet is a set of k prototype masks for the entire image. We set k = 32 to optimize speed and performance.

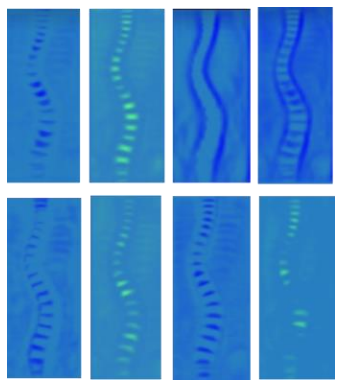

**Figure 4**: Sample Protonet Output. In each prototype, a different region of the image is activated. These prototypes will later be combined to produce the final segmentation.

### 3.1.2   Prediction Head

The second pathway is similar to RetinaNet, and it is used to predict mask coefficient. Usually, the prediction heads of anchor-based object detectors have two branches: one branch for *c* class confidences prediction, the other branch for bounding box regressors prediction [19]. YOLACT added a third branch in parallel for *k* mask coefficient prediction. Additionally, *tanh* is applied to the mask coefficients to produce more stable outputs over nonlinearity.

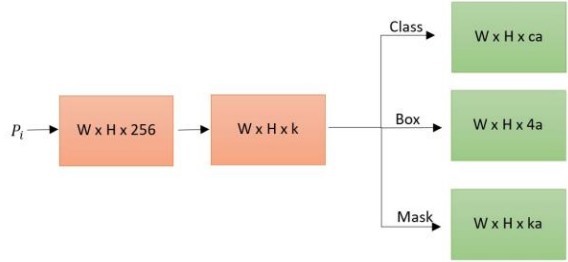

**Figure 5:** The Prediction Head Architecture of YOLACT has three branches. In addition to class confidences and bounding box regressors, it added a third branch for mask coefficient prediction.

## 3.2   Assembly

Lastly, YOLACT linearly combines the output of protonet and mask coefficient branch and applies sigmoid nonlinearity to produce the final masks. This is done through matrix multiplication and sigmoid function, according to Eq. 1, where P is a $w * h * k$ matrix from prototype masks and C is an $n * k$ matrix of mask coefficients.

$$M = \sigma(PC^T) \tag{1}$$

After mask assembly, three loss components were used to train the model: classification loss, box regression loss and mask loss with weights 1, 1.5, and 6.125, respectively. The classification loss and box regression loss are defined in the same way as the losses in Single Shot MultiBox Detector [20]. The mask loss is computed using pixel-wise binary cross-entropy between ground truth masks and the assembled masks.

## 3.3   Landmarks Extraction

Once we have the segmentation, we need to extract the landmarks of each vertebra. To do this, we first find the contour of each segmentation using opencv [21]. Then, we draw a minimum bounding box around the contour. The four corners of the minimum bounding box are considered to be the landmarks of the vertebra.

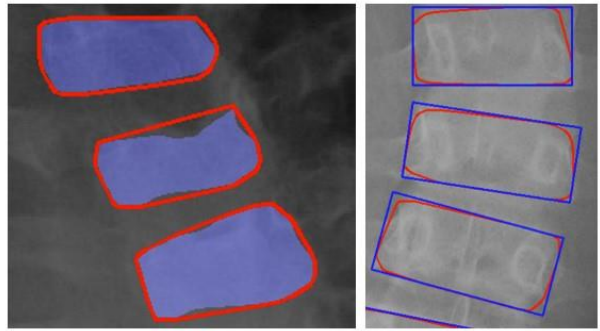

**Figure 6:** This figure demonstrates how a linear outline was used to identify the contour of vertebrae (a) initially, followed by drawing a minimum bounding box around the contour of each vertebra (b).

## 3.4  Angle Calculation

After correctly identifying the landmarks of each vertebra, we use trigonometry to calculate the Cobb angle. As shown in Figure 7, the strategy is to connect the dots to form a horizontally tilted line at each endplate [11]. We then develop an algorithm that iterates through all the lines to find three angles: 1) The maximum angle, which is usually in the Main-Thoracic (MT) region. 2) The maximum angle above MT, which is usually in the Proximal-Thoracic (PT) region, and 3)The maximum angle below MT, which is usually in the Thoraco-Lumbar(TL) region.

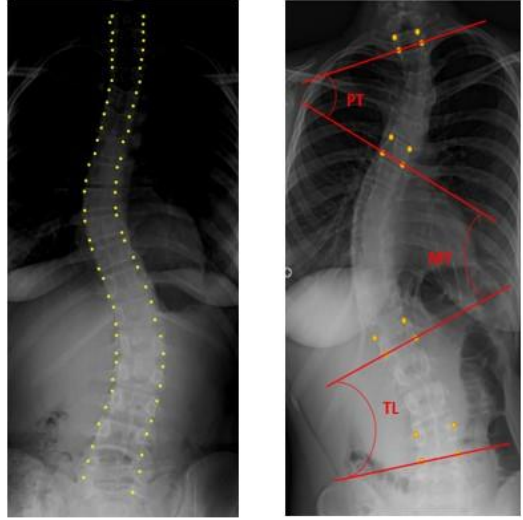

**Figure 7:** This figure demonstrates how to identify landmarks of individual vertebrae using dots (a) which are connected to form horizontally tilted lines at the lower vertebral endplates of the a priori selected landmark vertebrae resulting in three angles (b).

The angles between two lines can be calculated using trigonometry based on Eq.(2) where $m_1$ and $m_2$ are slopes for $l_1$ and $l_2$, respectively.

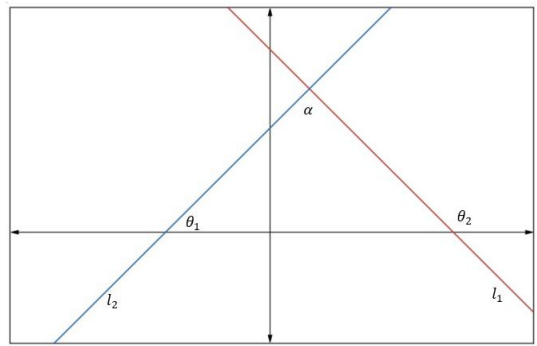

Figure 8: Calculating angle between two lines using Eq(2).

$$\begin{aligned} tan\alpha &= tan(\theta_1 - \theta_2) \\ &= \frac{tan\theta_1 - tan\theta_2}{1 + \theta_1) - tan\theta_2} \\ &= \frac{m_1 - m_2}{1 + m_1 m_2} \end{aligned} \qquad (2)$$

# 4 Experiment

## 4.1 Data Processing

The dataset of this research is obtained from SpineWeb [10], and includes 609 AP X-ray images in total. The ground-truth annotations are the anatomical landmarks consisting of four corners of 17 vertebrae: 12 thoracic and 5 lumbar. These landmarks were provided by two professional radiologists at London Health Science Center [22]. The dataset also contains the angle measurements of the proximal thoracic section, main thoracic section, and thoracolumbar section of the spine.

It should be noted that the input dataset contains flawed images. More specifically, it contains landmarks with incorrect ordering, as the accuracy of the landmark locations depends on the two clinicians' professional judgment. Figure 9 shows two examples of incorrect landmark ordering. We excluded the imperfect examples from the test set to avoid inaccuracies. The IDs of the excluded images can be found in Appendix A

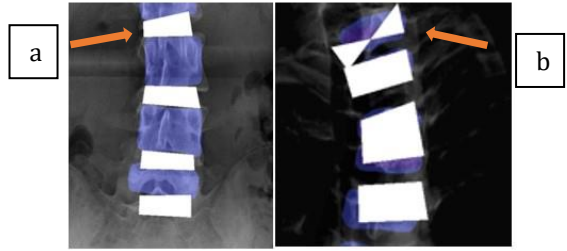

**Figure 9:** Examples of imperfect annotated data showing lateral vertebral wedging (arrow, a) and the shape of a "butterfly vertebra" (arrow, b).

## 4.2 Augmented Dataset

In order to expose the model to more diverse cases, we augmented the dataset using conventional methods. The data augmentation includes: 10% of the images are randomly tilted with an angle between -5 to 5 degrees; 10% of the images are flipped horizontally; 10% of the images are flipped vertically; and 10% of the images are normalized with histogram equalization.

To train the model, the images are split into training, validation, and testing sets with the 70%, 15%, and 15% percentages, respectively.

To input the images into YOLACT, COCO annotation format is required [18]. A custom COCO annotation contains detailed image information, such as the width and height of the image and the ground-truth location of the vertebra we want to segment [23]. Figure 10 illustrates an example input image and its COCO annotation. The quadrilaterals are created using the provided landmarks of the image.

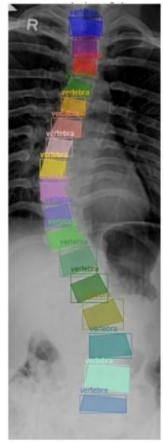

**Figure 10:** Coco Annotation displaying labeled vertebrae and bounding boxes. Each quadrilateral is created using the provided landmarks in the dataset

## 4.3 Evaluation Methods

With the calculated angles, the result of the proposed method can be evaluated using two methods. The first method is to use Symmetric Mean Absolute Percentage Error (SMAPE). SMAPE measures the relative error in the form of percentage.

$$SMAPE = \frac{1}{N} \sum_N \frac{\sum_{i=1}^{3}(|a_{ji}-b_{ji}|)}{\sum_{i=1}^{3}(|a_{ji}+b_{ji}|)} * 100\% \tag{3}$$

In Eq. 3, i indexes the three Cobb angles; j refers to the number of images; N is the total number of testing images; a is the estimated angles; and b is the ground-truth Cobb angles.

The second method is to use the absolute difference. The absolute difference can be calculated using Eq. 4 to gain more insight into the results.

$$\text{Absolute Difference} = |Groundtruth\ angle - predicted\ angle| \tag{4}$$

# 5 Results and Discussion

## 5.1 Qualitative result

Figure 11 illustrates the qualitative result of the YOLACT model. Overall, YOLACT produces acceptable segmentation. As shown, all segmentation results follow the curve of the spine with no major deviation.

It is observed that YOLACT tends to identify more vertebrae than desired. There are around 24 vertebrae in human spine; however, only 17 (12 thoracic vertebrae and 5 lumbar vertebrae) are used for Cobb angles measurement in this case [10]. YOLACT sometimes recognizes vertebra in the cervical region and produces segmentation for 18-19 vertebrae instead of 17. For example, in Figure 12, the white quadrilaterals indicate the ground truth, the blue segmentations are results from YOLACT. As shown, there is one extra vertebra identified at the top of the image.

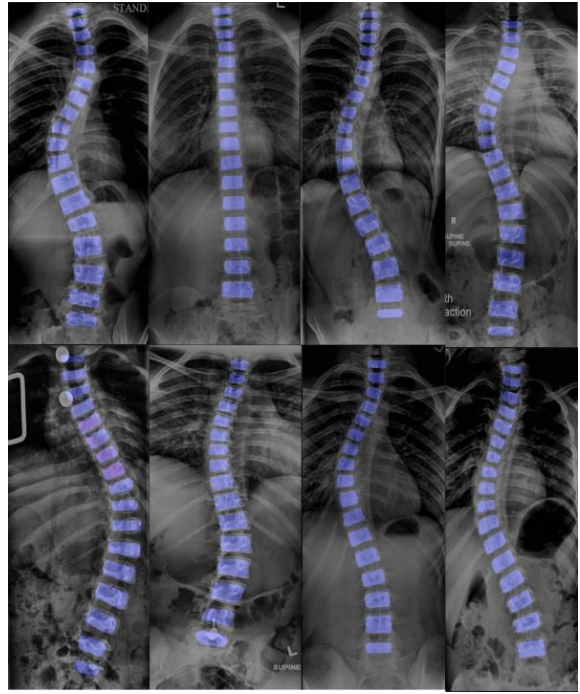

**Figure 11:** Qualitative Result of Spinal Vertebrae Segmentation. The blue segmentations are outputs of YOLACT. All segmentations follow the curve of the spine with no major deviation.

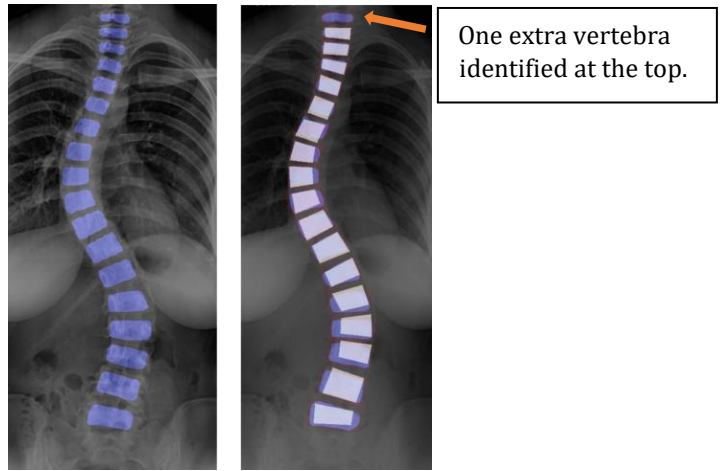

**Figure 12**: a) On the left is the segmentation result produced by YOLACT; b) On the right, we overlapped the segmentation ground truth (white quadrilaterals) on top of YOLACT's segmentation results (blue). As shown in the image, there is one extra blue vertebra identified at the top, as YOLACT sometimes recognizes vertebrae in the cervical region.

## 5.2 Quantitative result

Using Eq. 3, the calculated SMAPE for our approach is 10.76%, which outperforms the benchmarks discussed in the previous section. Table 1 lists the SMAPE results of the different approaches. The only model that has comparable performance with YOLACT is ResNet (SMAPE = 10.81%). Nevertheless, ResNet achieved the result by using two ML models instead of one. The ResNet approach first identifies the location of each vertebra; then, it finds the four corners using a trained center offset and heatmaps. L1 loss was used by the ResNet to train the corner offset [11]. The YOLACT approach uses only one model, and it has a SMAPE of 10.76%.

Table 1: SMAPE result comparison between different methods

| Methods | SMAPE |
|---|---|
| Revised U-Net | 16.48% |
| ResNet | 10.81% |
| Fast RCNN | 25.69% |
| Multi-View Extrapolation Net | 18.95% |
| $S^2VR$ | 37.08% |
| BoostNet | 23.44% |
| Fast RCNN | 25.69% |
| **YOLACT** | **10.76%** |

The absolute differences were also calculated using Eq. 4 to gain more insight into the results. The majority of the predictions had less than 5-degree differences, and 94.59% of predictions had error less than 10 degrees, indicating the reliability of the YOLACT model.

Table 2: Absolute angle difference between prediction and ground truth

| Absolute angle difference | Percentage |
|---|---|
| Difference less than 5 Degree | 64.86% |
| Difference between 5 and 10 degrees | 29.73% |
| Difference between 10 and 20 degrees | 5.41% |

# 6 Conclusion

In this paper, a modern instance segmentation model was used to estimate the Cobb angles over a dataset of 609 images. In comparison to previous methods, the YOLACT approach is accurate and consistent, and does not require an excessive amount of computation resources. We achieved a SMAPE of 10.76%, which demonstrates the ability of YOLACT to predict Cobb angles. To further improve on this method, one can implement outlier rejection techniques to correct segmentation error and remove undesirable noise as well as false positives objects.

# 7 Acknowledgment

This work was supported by Chair in Medical Imaging and Artificial Intelligence, a joint Hospital-University Chair between the University of Toronto, The Hospital for Sick Children, and the SickKids Foundation.

# 7 Appendix A

Table 3: Excluded images

| Image ID |
|---|
| sunhl-1th-01-Mar-2017-311 C AP.jpg |
| sunhl-1th-01-Mar-2017-312 C AP.jpg |
| sunhl-1th-14-Feb-2017-285 A AP.jpg |
| sunhl-1th-22-Feb-2017-291 E AP2.jpg |
| sunhl-1th-28-Feb-2017-243 C AP.jpg |
| sunhl-1th-28-Feb-2017-291 M AP.jpg |
| sunhl-1th-28-Feb-2017-291 N AP.jpg |
| sunhl-1th-28-Feb-2017-292 A AP.jpg |
| sunhl-1th-28-Feb-2017-292 B AP.jpg |
| sunhl-1th-28-Feb-2017-294 A AP.jpg |
| sunhl-1th-28-Feb-2017-295 I AP.jpg |